\documentclass[review]{elsarticle}
\usepackage{color}
\usepackage{amssymb}
\usepackage{lineno,hyperref}
\modulolinenumbers[5]

\journal{Journal of \LaTeX\ Templates}

\newcommand{\jpsi}{\mathrm{J/}\psi}

\newcommand{\WgPb}{W_{\gamma\mathrm{Pb}}}

\begin{document}

\begin{frontmatter}

\title{
Photonuclear $\mathbf{\mathrm{J/}\psi}$ production at the LHC: proton-based versus nuclear dipole scattering amplitudes
}

\author[CVUT]{D. Bendova}
\author[CVUT]{J. Cepila}
\author[CVUT]{J. G. Contreras}
\author[CVUT]{M. Matas}
\address[CVUT]{Faculty of Nuclear Sciences and Physical Engineering,
Czech Technical University in Prague, Czech Republic}


\begin{abstract}
The coherent photonuclear  production of a $\jpsi$ vector meson at the LHC has been computed using two different sets of solutions of the impact-parameter dependent Balitsky-Kovchegov equation. The nuclear dipole scattering amplitudes are obtained either from ($i$) solutions for this process off proton targets coupled with a Glauber-Gribov prescription, or ($ii$) from solutions obtained with an initial condition representing the nucleus. These approaches predict different cross sections, which are compared with existing data from ultra-peripheral collisions at the LHC.
The latter approach seems to better describe current measurements. Future LHC data should be precise enough to select one of the two approaches as the correct one.
\end{abstract}


\end{frontmatter}

\section{Introduction}

The exclusive photoproduction of a $\jpsi$ vector meson off a hadron has been recognised for many years as a very sensitive probe of the gluonic structure of hadrons in the perturbative regime of quantum chromodynamics (QCD)~\cite{Ryskin:1992ui,Brodsky:1994kf}; thus it has been extensively studied at HERA~\cite{Ivanov:2004ax,Newman:2013ada}. In recent years, this process has attracted renewed attention. On one hand, due to measurements at the LHC including production off protons and off Pb nuclei and reaching unprecedented energies~\cite{Baltz:2007kq,Contreras:2015dqa,Klein:2019qfb}. On the other, because of studies related to the potential of 
electron-ion colliders~\cite{Accardi:2012qut,AbelleiraFernandez:2012cc}.

 As mentioned above, there is plenty of high-quality data from HERA on production off proton targets. Therefore, many computations predicting the behaviour of this process off nuclear targets start from a description of the process off nucleons, where the parameters of the given model are fixed by HERA data, and then apply some form of Glauber formalism to predict the cross sections for photonuclear production. Such an approach has been followed for example in~\cite{Klein:1999qj,Lappi:2010dd,Goncalves:2014wna,Cepila:2017nef}.

The applicability of using a Glauber approach has been analysed since a long time, e.g.~\cite{Brodsky:1988xz,Kopeliovich:1991pu}, but recent advances in the understanding of saturation through the solution of the Balitsky-Kovchegov (BK) equation~\cite{Balitsky:1995ub,Kovchegov:1999yj,Kovchegov:1999ua} allow for new insights into this question.
In particular, the implementation of collinear corrections to the kernel~\cite{Iancu:2015vea,Iancu:2015joa}
together with a suitable initial condition have been used to find impact-parameter dependent solutions of the BK equation~\cite{Cepila:2018faq}, which correctly describe HERA data on vector meson photo- and electroproduction off protons~\cite{Bendova:2019psy}. 

Recently, these advances have been extended to the case of nuclear targets~\cite{Cepila:2020xol} using two approaches: ($i$) coupling the solution of the BK equation for the case of proton targets to a Glauber-Gribov prescription to obtain the solutions to the nuclear case, and ($ii$) solving directly the impact-parameter dependent BK equation with an initial condition representing a specific nucleus. In what follows, these two set of solutions are denoted as b-BK-GG and b-BK-A, respectively.

In this Letter, both approaches are used to predict the cross section for coherent photoproduction of $\jpsi$ vector mesons in Pb--Pb ultra-peripheral collisions (UPC) at the LHC and compare the predictions with data available at different rapidities and at two centre-of-mass energies per nucleon pair, $\sqrt{s_{\rm NN}} = 2.76$~TeV and $\sqrt{s_{\rm NN}} = 5.02$~TeV, corresponding to measurements performed during the LHC Run 1 and Run 2, respectively. It is found that Run 1 measurements at midrapidity strongly disfavour the use of b-BK-GG solutions, and that the expected precision of the measurements with Run 2 data may provide a definitive answer on the question of which approach is the valid one. The rest of this Letter is organised as follows: the next section presents a brief overview of the formalism; Sec.~\ref{sec:Results} contains the main results, while in Sec.~\ref{sec:Discussion} our findings are discussed; the Letter concludes with a brief summary and outlook in Sec.~\ref{sec:Summary}.
 
\section{Brief overview of the formalism
\label{sec:formalism}}
In this section a brief overview of the formalism is presented. For the full details see for example~\cite{Bendova:2019psy,Cepila:2020xol} and references therein.

The cross section for the coherent photoproduction of a $\jpsi$ vector meson, differential on the square of the momentum transfer $t$ at the target vertex, is given by the sum of the contributions from transversely ($T$) and longitudinally ($L$) polarised photons:
\begin{equation}
\frac{\mathrm{d}\sigma_{\gamma{\rm Pb}}}{\mathrm{d}|t|} \bigg| _{T,L} = \frac{\left(1+\beta^2\right) \left(R_g ^{T,L}\right)^2}{16\pi} | \mathcal{A}_{T,L} |^2.
\label{VM-cs-diff-excl}
\end{equation}
 The factor $(1+\beta^2)$ accounts for contributions from the real part of the amplitude, while $(R_g ^{T,L})^2$ corrects for the so-called skewedness~\cite{Shuvaev:1999ce}. The scattering amplitude of the process is given by
\begin{equation}
\mathcal{A}_{T,L}(x,Q^2,\vec{\Delta}) = i \int \mathrm{d}\vec{r} \int \limits_0^1 \frac{\mathrm{d}z}{4\pi} \int \mathrm{d}\vec{b} |\Psi_{\rm V}^* \Psi_{\gamma^*}|_{T,L} \exp \left[ -i\left( \vec{b} - (1-z)\vec{r} \right)\vec{\Delta} \right] \frac{\mathrm{d}\sigma^{q\bar{q}}}{\mathrm{d} \vec{b}}.
\label{VM-amplitude}
\end{equation}
Here, $\Psi_{\gamma^*}$ and $\Psi_{\mathrm{V}}$ are the wave functions of a virtual photon fluctuating into a colour dipole and of the dipole producing the vector meson. The vector $\vec{r}$ represents the dipole size and orientation, and  $\vec{b}$ represents the impact parameter between the dipole and the target. $Q^2$ denotes the virtuality of the photon and $\vec{\Delta}^2 \equiv -t$. The variable $z$ corresponds to the fraction of the energy of the quark-antiquark dipole carried by the quark, while
\begin{equation}
\frac{\mathrm{d}\sigma^{q\bar{q}}}{\mathrm{d} \vec{b}} = 2N(\vec{r},\vec{b};x),
\end{equation}
with $N(\vec{r},\vec{b};x)$ the dipole scattering amplitude obtained as a solution of the BK equation at a rapidity $Y=\ln(x_0/x)$; here $x_0\equiv 0.008$ corresponds to the rapidity at the initial condition. 

As mentioned before, two sets of dipole scattering amplitudes are used; both were obtained and studied in detail in our previous work~\cite{Cepila:2020xol}\footnote{The amplitudes are available online at \url{https://hep.fjfi.cvut.cz/NuclearbdepBK.php}}.
In the b-BK-GG approach, the impact-parameter dependent BK equation is solved with an initial condition representing a proton. The solutions at each rapidity are then converted into solutions for a nucleus using the  
Glauber-Gribov prescription proposed in~\cite{Armesto:2002ny}. In the case of the b-BK-A approach, the initial condition represents the specific nucleus where the impact-parameter part is described with the help of the corresponding Woods-Saxon distribution.

\section{Results
\label{sec:Results}}
\begin{figure}[t!]
\centering
\includegraphics[width=0.48\textwidth]{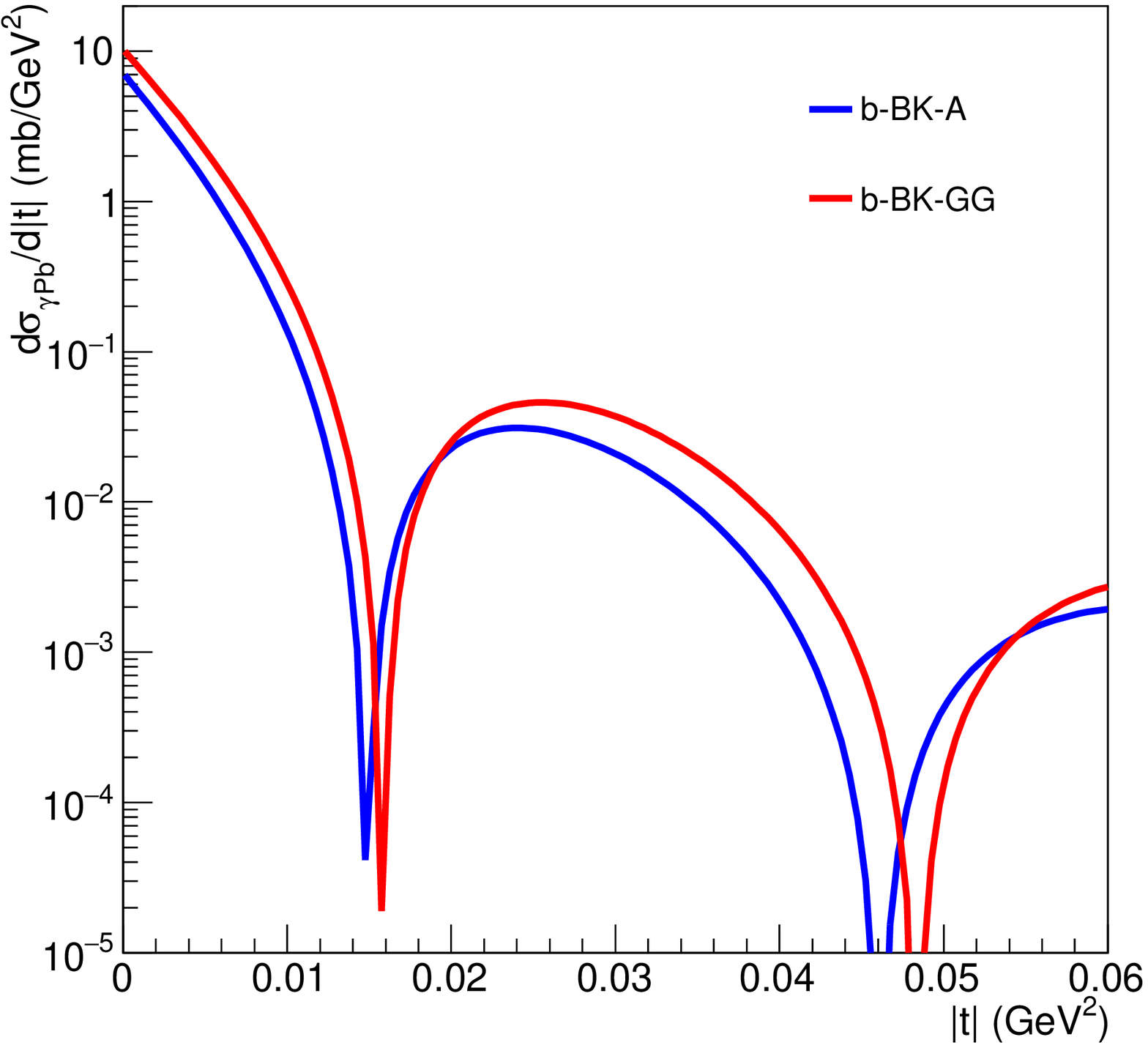}
\includegraphics[width=0.48\textwidth]{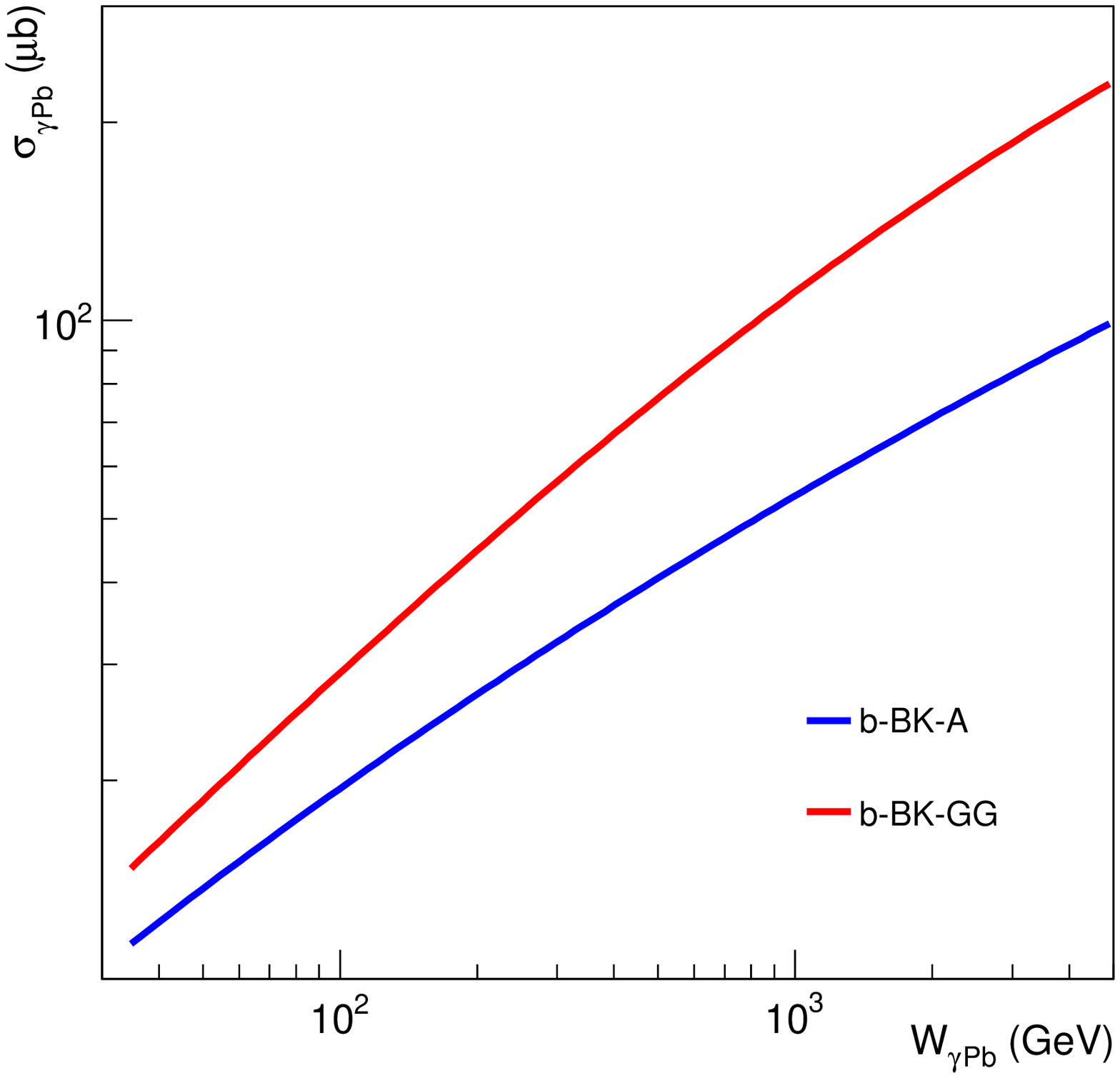}
\caption{\label{fig:t_dist} (Colour online) Left: Cross section for the coherent photoproduction of a $\jpsi$ vector meson off a Pb target as a function of $|t|$ at a centre-of-mass energy of the $\gamma{\rm Pb}$ system $W_{\gamma{\rm Pb}} = 121$ GeV. Right: Energy dependence for the cross section integrated over $|t|$.}
\end{figure}

The cross section for the coherent photoproduction of a $\jpsi$ vector meson off a Pb target as a function of $|t|$ is shown in Fig.~\ref{fig:t_dist} (left) at a centre-of-mass energy of the $\gamma{\rm Pb}$ system $W_{\gamma{\rm Pb}} = 121$ GeV, where $W^2_{\gamma{\rm Pb}} =M^2_{\jpsi}/x$ with $M_{\jpsi}$ the mass of the $\jpsi$ vector meson. Note that not only the absolute magnitude of the cross section is different in the b-BK-A and b-BK-GG approaches, but also that the positions of the diffractive minima are displaced. This particular value of $W_{\gamma{\rm Pb}}$ has been chosen, because it corresponds to production in UPC at midrapidity for LHC Run 2 $\sqrt{s_{\rm NN}}$ energies, as explained below.

\begin{figure}[t!]
\centering
\includegraphics[width=0.48\textwidth]{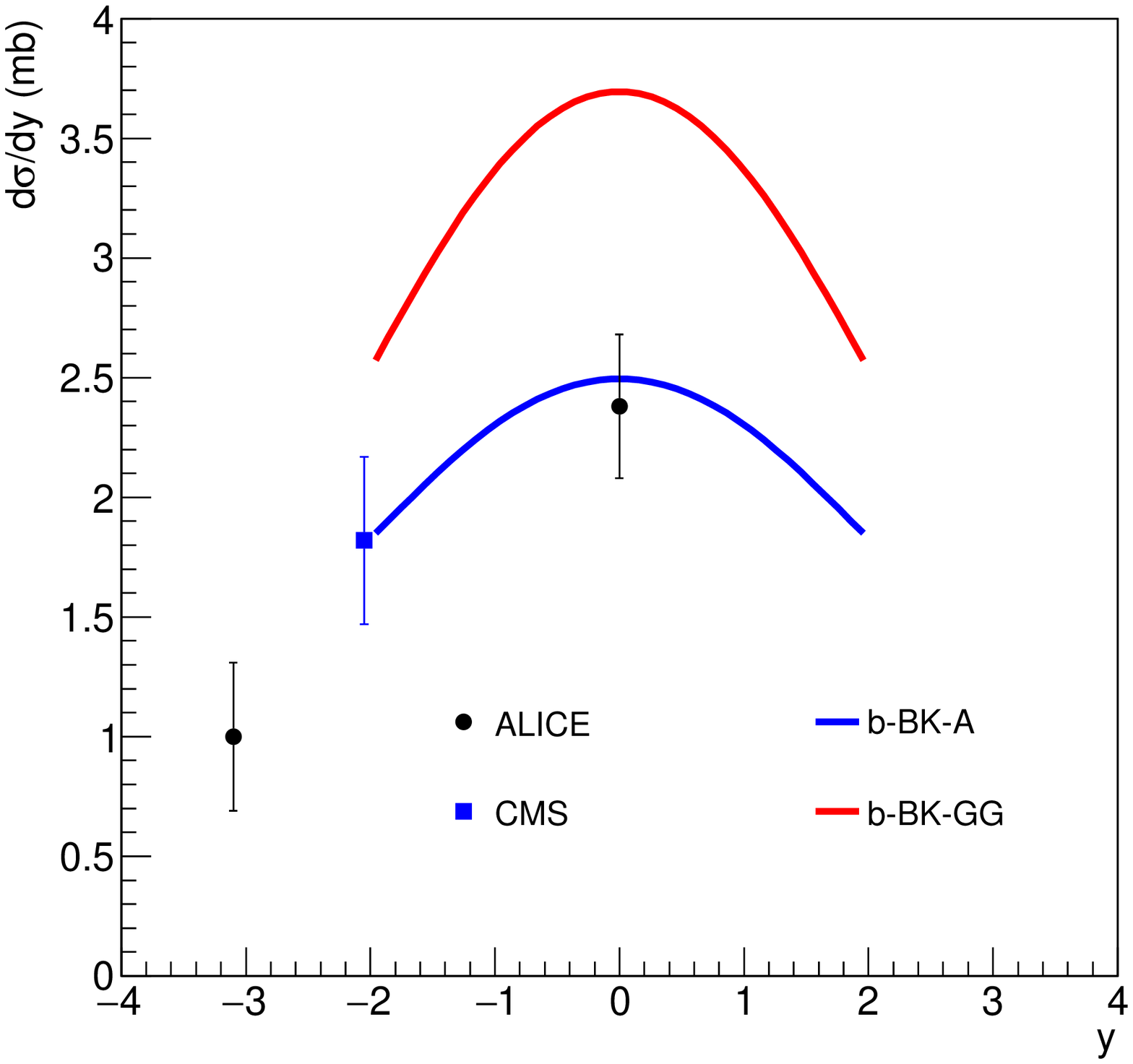}
\includegraphics[width=0.48\textwidth]{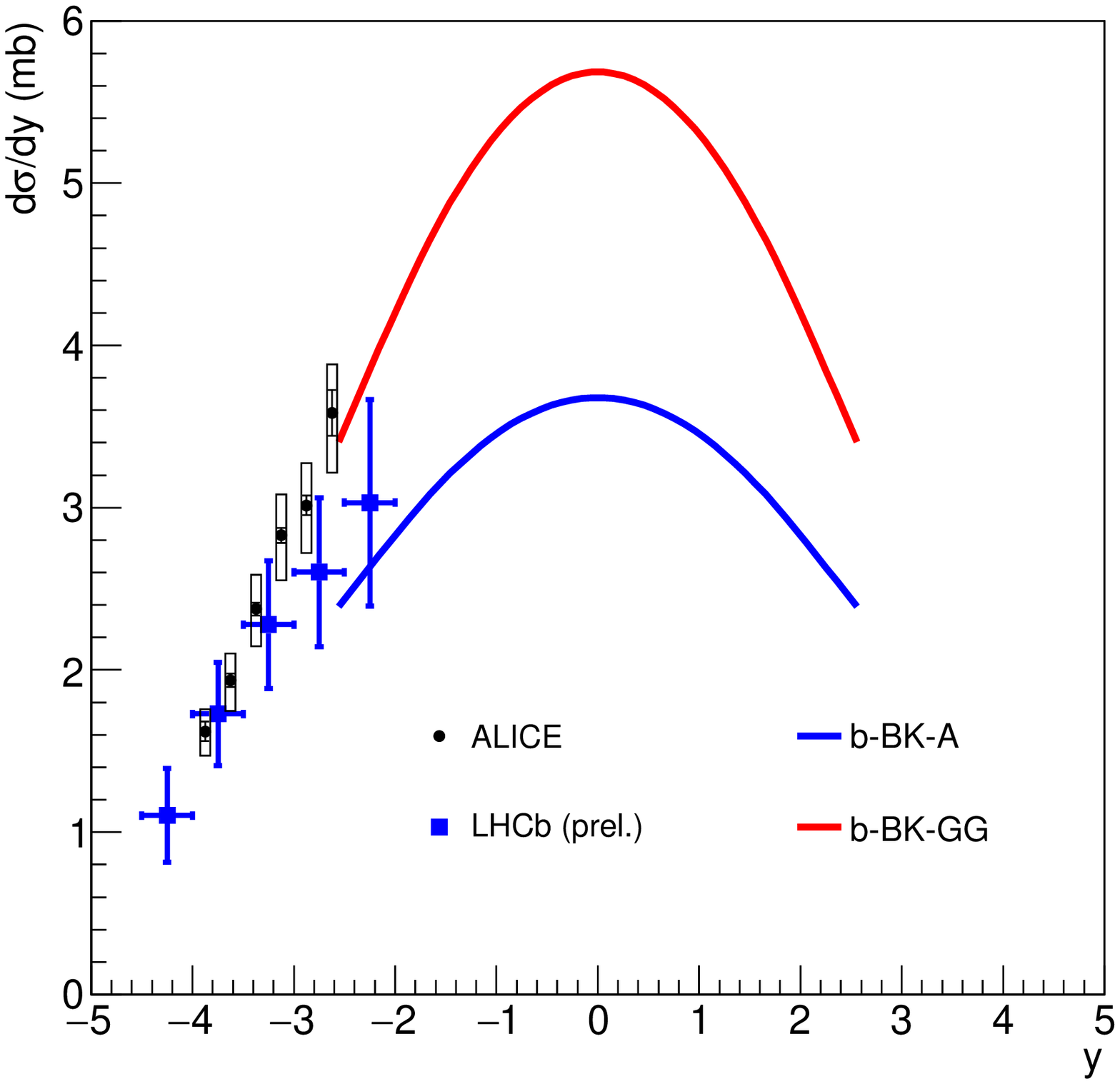}
\caption{\label{fig:run12} (Colour online) Cross section for the coherent photoproduction of a $\jpsi$ vector meson in ultra-peripheral Pb--Pb collisions at $\sqrt{s_{\rm NN}} = 2.76$~TeV (left) and $\sqrt{s_{\rm NN}} = 5.02$~TeV (right) corresponding to LHC energies during the the Run 1 and Run 2 periods, respectively. The predictions are compared with data from the ALICE~\cite{Abelev:2012ba,Abbas:2013oua,Acharya:2019vlb} and CMS~\cite{Khachatryan:2016qhq} collaborations as well as with preliminary results from the LHCb collaboration~\cite{LHCb:2018ofh}.}
\end{figure}

Figure~\ref{fig:t_dist} (right) shows the energy dependence of the total $\gamma{\rm Pb}$ cross section, that is integrated over $|t|$. The difference in the absolute value of the cross section when using b-BK-A with respect to b-BK-GG solutions increases with energy from a 30\% at $W_{\gamma{\rm Pb}} = 35$ GeV to 54\% at $W_{\gamma{\rm Pb}} = 121$ GeV, reaching already a factor of two at $W_{\gamma{\rm Pb}} = 900$ GeV. 

The cross section $d\sigma/dy$ for the coherent photoproduction of a $\jpsi$ vector meson in Pb--Pb UPC is shown in Fig.~\ref{fig:run12} for the LHC energies corresponding to the Run 1 (left) and Run 2 (right). This cross section is given by
\begin{equation}
\frac{d\sigma}{dy} = 
n_{\gamma}(y)\sigma_{\gamma \rm Pb}(y)+
n_{\gamma}(-y)\sigma_{\gamma \rm Pb}(-y), 
\label{eq:XS}
\end{equation}
where the rapidity $y$ of the $\jpsi$ at the LHC is related to $W_{\gamma{\rm Pb}}$ by
\begin{equation}
\WgPb^2 = \sqrt{s_{\rm NN}}M_{\jpsi}e^{-y}.
\label{eq:WsNNy}
\end{equation}
The flux of photons from the Pb nucleus $n_{\gamma}(y)$ is computed following the description detailed in~\cite{Contreras:2016pkc}. The figure also shows a comparison with existing measurements from the ALICE~\cite{Abelev:2012ba,Abbas:2013oua,Acharya:2019vlb} and CMS~\cite{Khachatryan:2016qhq} collaborations as well as with preliminary results from the LHCb collaboration~\cite{LHCb:2018ofh}.

\section{Discussion
\label{sec:Discussion}}

Some comments are in order. First, those of technical nature are addressed, followed by those related to the physics insight provided by the results presented in the previous section.

There has been  recent interest on the argument of the exponential term in Eq.~(\ref{VM-amplitude}). This factor, introduced in~\cite{Bartels:2003yj}, originates from a Fourier Transform term modified to take into account non-forward amplitudes. In~\cite{Bartels:2003yj} the factor is written in a general form, but when used for phenomenology it has been commonly implemented as in Eq.~(\ref{VM-amplitude}). A proposal put forward in~\cite{Hatta:2017cte} and based on symmetry arguments is that the term $(1-z)$ should be $(1-2z)/2$. 
Using the proposal from~\cite{Hatta:2017cte} produces a 3.5\% larger cross section in both the b-BK-A and the b-BK-GG scenarios. This percentage is constant within the studied energy range. Therefore, this issue does not affect significantly the results presented in this Letter.

The corrections to take into account contributions from the real part of the amplitude and the skewedness effect are computed at fixed $|t|=0.0001$. They depend on energy decreasing slowly with increasing $\WgPb$. The factor $(1+\beta^2)$ is 1.07 (1.08) around 35 GeV and 1.04 (1.05) at 1 TeV, while $(R_g ^{T,L})^2$ is 1.32 (1.34) around 35 GeV and 1.23 (1.27) at 1 TeV for the b-BK-A (b-BK-GG) case.

The predictions shown in Fig.~\ref{fig:run12} cover a restricted range in rapidity. The origin of this limitation is that the initial condition for the evolution of the dipole scattering amplitude in the BK equation corresponds to an initial value of $x_0=0.008$. Inserting this into $W^2_{\gamma{\rm Pb}} =M^2_{\jpsi}/x$ and using Eq.~(\ref{eq:WsNNy}) produces a lower limit in $y$ for Eq.~(\ref{eq:XS}).

The approach followed here to compare the predictions from the b-BK-A and b-BK-GG is consistent in the sense that the same wave functions and the same corrections are used. The internal parameters not directly related to the targets take the same values in both cases and the subjacent QCD input, namely the BK equation with the collinear corrections, is the same. Furthermore, this implementation of the BK equation and the corresponding solutions including the impact-parameter dependence avoid the introduction of ad hoc parameters or assumptions to describe the distribution of matter in the plane transverse to the $\gamma A$ interaction. The solutions for the proton case used in the b-BK-GG approach described correctly  photo and electroproduction data from HERA \cite{Bendova:2019psy}.

The cross sections shown in Fig.~\ref{fig:t_dist} (left) demonstrate the presence of diffractive dips. The location of the dips have been put forward as a signature of saturation in $\gamma$p~\cite{Armesto:2014sma} and $\gamma A$ collision~\cite{Toll:2012mb}. The facts that the position of the dip changes according to whether a Glauber-Gribov prescription is used or not, and that the change is larger than that observed in~\cite{Toll:2012mb} between the saturation and the no-saturation cases, casts a warning on the use of this observable. 

The flux entering Eq.~(\ref{eq:XS}) is fairly constant for lower $\WgPb$ energies, but it shows a strong cut-off at large energies. As the $\gamma A$ cross section raises with energy as shown in Fig.~\ref{fig:t_dist} (right), the two terms in Eq.~(\ref{eq:XS}) have a different numerical value at large $|y|$ with the low $\WgPb$ contribution being dominant. In this region, the predictions for the b-BK-A and b-BK-GG prescriptions are the closest. At midrapidity, both contributions to Eq.~(\ref{eq:XS}) are the same and correspond to $\WgPb=125$ GeV. Here, the difference in the presented UPC cross sections is the largest as shown in Fig.~\ref{fig:run12}. Comparison with data from the LHC Run 1 indicates a preference for the b-BK-A approach and disagrees with b-BK-GG at a bit more than one-sigma for $|y|=2$ and more than 3 sigmas for $y=0$. The currently existing data from the LHC Run 2 does not provide such a clean message because of the large experimental uncertainties as well as the slight apparent discrepancy between ALICE and LHCb results. The data from LHC Run 2 at midrapidity are still being analysed; it is expected that the uncertainties will be smaller than those in the existing measurement. If so, then these new data may help to select one of the two prescriptions as the most adequate approach.

These results, specifically those shown in Fig.~\ref{fig:t_dist}, are of interest for future electron-ion colliders~\cite{Accardi:2012qut,AbelleiraFernandez:2012cc} where such a process will be precisely measured for a variety of nuclei, allowing for the study not only of the energy, but also of the $A$ dependence of the cross section for coherent $\jpsi$ photo and electroproduction.

\section{Summary and outlook
\label{sec:Summary}}
The coherent photonuclear production off Pb nuclei in ultra-peripheral collisions at the LHC has been studied using solutions of the impact-parameter dependent BK equation. Two approaches have been compared. Starting from solutions of the proton case coupled to a Glauber-Gribov formalism, or solving directly the impact-parameter dependent BK equation with an initial condition representing the nucleus.
Data from the LHC favour the latter approach. Future data at midrapidity should be precise enough to settle the question of the most valid approach in this context. These studies are of interest for the newly approved and planned future electron-ion colliders where this type of process can be studied with  more precision and in a variety of ways.

\section*{Acknowledgements}
This work has been partially supported by  grant 18-07880S of the Czech Science Foundation (GACR), grant LTC17038 of the INTER-EXCELLENCE program at the Ministry of Education, Youth and Sports of the Czech Republic and by the Centre of Advanced Applied Sciences with the number: CZ.02.1.01/0.0/0.0/16-019/0000778. The Centre of Advanced Applied Sciences is co-financed by the European Union. 

\bibliography{b-BK-VM}

\end{document}